\documentclass[journal=jacsat,manuscript=article]{achemso}

\usepackage[
]{achemso}
\setkeys{acs}{maxauthors=10}
\setkeys{acs}{etalmode=truncate}
\usepackage[version=3]{mhchem} 
\usepackage{esint}
\usepackage{threeparttable}
\usepackage{xr-hyper} 
\usepackage{hyperref} 
\usepackage{cleveref}
\usepackage[dvipsnames]{xcolor}
\usepackage{float}
\usepackage{caption}
\mciteErrorOnUnknownfalse

\makeatletter
\newcommand*{\addFileDependency}[1]{
  \typeout{(#1)}
  \@addtofilelist{#1}
  \IfFileExists{#1}{}{\typeout{No file #1.}}
}
\makeatother

\makeatletter
\newcommand*{\addAuxFileDependency}[1]{
  \makeatletter\@input{x#1.tex}\makeatother
}
\makeatother

\newcommand*{\myexternaldocument}[1]{%
    \externaldocument[#1:]{#1}%
    \addAuxFileDependency{#1}
    \addFileDependency{#1.tex}%
    \addFileDependency{#1.aux}%
}

\author{Sushree Jagriti Sahoo}
\affiliation[Georgia Institute of Technology]
{School of Chemical and Biomolecular Engineering, Georgia Institute of Technology, Atlanta, Georgia 30318 USA}

\author{Xin Jing}
\affiliation[Georgia Institute of Technology]
{School of Computational Science and Engineering, Georgia Institute of Technology, Atlanta Georgia 30313 USA} 
\author{Phanish Suryanarayana}
\affiliation[Georgia Institute of Technology]
{School of Civil and Environmental Engineering, Georgia Institute of Technology, Atlanta Georgia 30332 USA}
\author{Andrew J. Medford}
\affiliation[Georgia Institute of Technology]
{School of Chemical and Biomolecular Engineering, Georgia Institute of Technology, Atlanta, Georgia 30318 USA}
\email{ajm@gatech.edu}

\crefname{equation}{Eq.}{Eqs.}
\Crefname{equation}{Equation}{Equations}
\crefname{table}{Table}{Tables}
\Crefname{table}{Table}{Tables}
\crefname{figure}{Fig.}{Figs.}
\Crefname{figure}{Figure}{Figures}

\title{Ab-initio investigation of finite size effects in rutile titania nanoparticles with semilocal and nonlocal density functionals}

\abbreviations{IR,NMR,UV}

\begin{document}
\myexternaldocument{supp}






\newpage
\begin{abstract}
In this work, we employ hybrid and generalized gradient approximation (GGA) level density functional theory (DFT) calculations to investigate the convergence of surface properties and electronic gap of rutile titania (TiO$_2$) nanoparticles with particle size. The surface energies and electronic gaps are calculated for cuboidal particles with minimum dimension ranging from 3.7 \AA{} (24 atoms) to 10.3 \AA{} (384 atoms) using a highly-parallel real-space DFT code to enable hybrid level DFT calculations of larger nanoparticles than are typically practical.  We deconvolute the geometric and electronic finite size effects in surface energy, and evaluate the influence of defects on electronic gap and density of states (DOS). The electronic finite size effects in surface energy vanish when the minimum length scale of the nanoparticles becomes greater than 10 \AA{}. We show that this length scale is consistent with a computationally efficient numerical analysis of the characteristic length scale of electronic interactions. The surface energy of nanoparticles having minimum dimension beyond this characteristic length can be approximated using slab calculations that account for the geometric defects. In contrast, the finite size effects on the electronic gap and DOS is highly dependent on the shape and size of these particles. The DOS for cuboidal particles and more realistic particles constructed using the Wulff algorithm reveal that defect states within the electronic gap play a key role in determining the eigen value distribution of nanoparticles and the electronic gap does not converge to the bulk limit for the particle sizes investigated. 
\end{abstract}

\newpage
\section{Introduction}
Metal oxide nanoparticles have wide range of commercial and technological applications ranging from biomedical engineering to chemical catalysis.\cite{Bourikas2014TitaniumCatalysts,siriwong2012doped,alexandrov2020ga,DIZAJ2014278,parham2016antimicrobial,yang2014stable,pal2015faceted} Oxide nanoparticles have a number of favorable properties including a high surface-area to volume ratio, tunable optical properties, and a range of surface reactivities. For example, oxide nanoparticles can be used as inert support materials in catalysis,\cite{kamat1993photochemistry} or provide reactive surfaces that interact with supported catalysts,\cite{zhou2008nanoparticles} or act directly as catalysts.\cite{kamat1993photochemistry,JIANG199824} In particular, oxide nanoparticles are commonly used for photocatalysis since it requires high surface areas, electronic gaps that are aligned with the redox potentials of products and reactants, and surface chemistry that facilitates chemisorption and reaction of intermediates.\cite{walter2010solar,osterloh2013inorganic} Both the electronic gap and chemical reactivity of nanoparticles are controlled by their atomic structure and size, as well as the surrounding environment such as solvents, capping agents and supports. However, atomic-scale simulations of nanoparticles are challenging due to the number of atoms and electrons present, especially if their environment is considered. Therefore, it is a natural starting point to study the structure and reactivity of isolated nanoparticles to establish an understanding of the key factors governing the electronic gap and reactivity to help design nanoparticles with specific optical and catalytic properties.\cite{C3CS60421G} 
 
 In particular, TiO$_2$ is of paramount technological importance. The applications of TiO$_2$ include hydrogen production, photo-voltaic cells, degradation of harmful organic compounds in the environment, medical applications such as bone implants, and supports for other catalytic materials.\cite{Bourikas2014TitaniumCatalysts,AUGUSTINE201791,azam2012antimicrobial,zhang2004synthesis,nakajima2014selective,MAIRA2001327} For this reason, TiO$_2$ is widely studied in the scientific literature and is often referred to as a ``model oxide''.\cite{Fujishima2000TitaniumPhotocatalysisb,DIEBOLD200353,Carp2004PhotoinducedDioxideb} A large number of ab-initio studies focus on models of bulk polymorphs,\cite{Mattioli2008iAbClass,Finazzi2008ExcessCalculations,Mattioli2010DeepPolymorphs,S.Bjrheim2010Ab2} extended surfaces\cite{Lazzeri2001StructureSurfaces,Linsebigler1995PhotocatalysisResults,Morgan2007ASurface} and small nanoparticles but relatively few studies have been done on investigation of finite size effects arising in small TiO$_2$ nanoparticles,\cite{Lamiel-Garcia2017WhenCalculationsb,Selli2017ModellingDFT,Auvinen2011SizeNanoparticles} especially for the rutile polymorph of TiO$_2$. These finite size effects can be classified into geometric effects which arise due to unique atomic configurations in the nanoparticle, electronic effects, which arise due to features in the electronic structure of the nanoparticle system, and quantum effects, which arise due to quantization of energy levels in small particles.\cite{Li2013InvestigationClusters} Since rutile TiO$_2$ has a large number of applications in the field of heterogeneous catalysis and photocatalysis,\cite{Clawin2014DefectsTiO2110,Kipreos2018WaterDRIFTS,Nwankwo2019SynthesisApplications} it represents a natural starting point for evaluating the finite size effects on surface and electronic properties. In this work, we study the finite size effects in surface energy which is an important property that dictates the surface structure and stability. The electronic property in consideration is the electronic gap, which is crucial in applications of photocatalysis because nanoparticles of TiO$_2$ are often used, but the interplay between nanoparticle structure and photon absorption is not well understood.   

Previous works in the literature have investigated the intrinsic particle size effect with increasing size of metallic nanoparticles using adsorption energy to characterize surface catalytic properties. \citeauthor{Kleis2011FiniteSolids}\cite{Kleis2011FiniteSolids} performed a DFT study using the revised Perdew-Burke-Ernzerhof (RPBE) functional\cite{Hammer1999ImprovedFunctionals} on gold metal nanoparticles ranging from 13 to 1,415 atoms to show how surface properties varied with system size at two local geometries that resemble surfaces of (111) and (211) slabs. They show that surface properties converge to the slab limit at a characteristic length of 27 \AA{} (560 atoms). The generality of findings for other transition metals were confirmed with a similar study on freestanding cuboctahedral platinum metal nanoparticles which show analogous convergence with size but at a smaller characteristic length of 16 \AA{} (147 atoms).\cite{Li2013InvestigationClusters} The main limitation of the prior work described above is that they focus on metallic systems and are limited to semilocal GGA exchange-correlation functionals. Exceptions include several studies focused on finite size effects in anatase\cite{Lamiel-Garcia2017WhenCalculationsb, morales2019understanding} and rutile \cite{ko2017size} nanoparticles at the hybrid level of theory with light-tier-1 numerical atom-centered orbitals basis. However, the convergence of surface properties toward the slab limit was not investigated, and the particles were constructed to minimize the presence of geometric defects which may play a significant role in catalytic systems.

\citeauthor{Jinnouchi2017PredictingAlgorithm}\cite{Jinnouchi2017PredictingAlgorithm} have also shown that metal alloy nanoparticles can contain heterogeneous atomic configurations such as atomic-scale defects that cannot be explained by the single-crystal surface models, and that these defects dominate the catalytic activity of nanoparticles. They propose a machine-learning scheme which adopts a descriptor-based approach to map slab-based surface models to the nanoparticle surfaces and defects. The machine-learning model is trained on single crystal slabs with various defects and compositions. The model is then able to accurately predict energetics of complex nanoparticles, which indicates that the geometric defects dominate the nanoparticle behavior. However, it is unclear if a similar strategy will work for (mixed) metal oxide particles, particularly for photocatalysts where the electronic structure is important.


Evaluating finite size effects on electronic structure properties is complicated by the need for hybrid-level functionals. Exchange-correlation functionals based on the local-density approximation (LDA) and GGA significantly underestimate the electronic gap because of a derivative discontinuity in the exchange-correlation potential.\cite{Borlido2020Exchange-correlationLearning} However, the hybrid functionals that are implemented within the generalized Kohn-Sham (KS) formalism instead of standard KS formalism incorporate part of the discontinuity which leads to a electronic gap that is in good agreement with experimental values.\cite{PhysRevB.53.3764} Despite the fact that planewave codes are the most widely used method for solving the KS equations, it becomes impractical to perform planewave hybrid calculations on systems with $>$ 100 atoms due to the early onset of the cubic scaling bottleneck and the large associated prefactor. In particular, the limited scalability on parallel computing platforms restricts the time to solution that can be achieved.  Another disadvantage of planewave codes is the nature of the Fourier basis that restricts the method to periodic boundary conditions, hence making it necessary to add artificial periodicity using vacuum and dipole corrections, which limits accuracy in the study of systems with dipole moments as well as charged systems.\cite{XU2021100709,PhysRevB.73.115407,PhysRevB.51.4014} The use of a finite difference basis set can overcome these challenges by enabling ideal parallelization and application of non-periodic boundary conditions.


In this work, we investigate finite size effects in rutile TiO$_2$, which is the most stable polymorph of titania.\cite{Bagayoko2011ToBagayoko} We utilize a collection of model systems including bulk, surface slabs, and nanoparticles to deconvolute effects arising from various types of defects. We use these model systems to elucidate finite size effects on the surface energy and electronic gap, and we explore the influence of the exchange correlation functional on these effects by performing calculations with both semilocal (PBE) and hybrid (PBE0) functionals. The study utilizes the new finite-difference Simulation Package for Ab-initio Real-space Calculations (SPARC)\cite{XU2021100709} code to enable hybrid-level calculations on nanoparticle systems. We also utilize a numerical convergence technique to assess the length scale of electronic interactions, and apply a geometric fingerprinting scheme based on machine-learning\cite{PhysRevLett.98.146401} to deconvolute geometrical and electronic finite size effects\cite{Li2013InvestigationClusters} in surface energy. The results indicate that the surface energy converges quickly for both functionals, approaching the infinite particle limit at particle sizes of $\sim$ 10 \AA{}. However, we find that the electronic gap and DOS are highly sensitive to functional choice, particle size, and particle shape, indicating that hybrid-level calculations are required to assess the electronic structure of nanoparticles well beyond the sizes investigated here.

\section{Methods}

\subsection*{Kohn-Sham DFT simulations}\label{subsec:dft_simulations}
We perform Kohn-Sham DFT calculations using the state-of-the-art ``Simulation Package for Ab-Initio Real-space Calculations'' (SPARC) software \cite{XU2021100709, ghosh2017sparc2, ghosh2017sparc1} --- a real-space DFT code that has comparable accuracy to established planewave codes, while requiring walltimes that are more than an order of magnitude lower. In all calculations, we neglect spin and choose optimized norm-conserving Vanderbilt (ONCV) pseudopotentials \cite{Hamann2013OptimizedPseudopotentials} from the SG15 \cite{SCHLIPF201536} collection. In addition, we employ the Perdew-Burke-Ernzerhof (PBE) \cite{Perdew1996GeneralizedSimple} and PBE0 \cite{perdew1996rationale, adamo1999toward} exchange-correlation functionals for performing  GGA and hybrid level calculations, respectively.  

In all simulations, we choose the twelfth-order finite-difference approximation. For bulk calculations, where periodic boundary conditions are prescribed in all three coordinate directions, we employ a mesh-size of 0.25 bohr and 4 $\times$ 4 $\times$ 4 Monkhorst-Pack grid \cite{monkhorst1976special} for Brillouin zone integration. These and other parameters have been chosen to provide an accuracy of 0.01 bohr in the computed equilibrium lattice constants. For slab calculations, where periodic and Dirichlet boundary conditions are prescribed in the plane and perpendicular to the plane of the slab, respectively, we employ a mesh-size of 0.25 bohr, a 4 $\times$ 4  Monkhorst-Pack grid for Brillouin zone integration, and a vacuum of 8 bohr. These and other parameters have been chosen to provide an accuracy of 0.001 Ha/atom in the energy. For the nanoparticle calculations, where Dirichlet boundary conditions are employed in all three coordinate directions, we employ a mesh-size of 0.3 bohr and vacuum of 8 bohr in each coordinate direction. These and other parameters have been chosen to provide an accuracy of 0.001 Ha/atom in the energy. Note that we perform structural relaxation for only the bulk system, while the atoms in the slab systems and nanoparticles are held fixed, i.e., only the electronic ground state is computed for the given atomic positions. All atomic positions and structures are defined using the Atomic Simulations Environment (ASE) package.\cite{HjorthLarsen2017TheAtomsb}

\subsection*{Linear regression model for nanoparticle energetics}
We adopt the following decomposition for the nanoparticle energy:
\begin{equation} \label{eq:total_energy_reg}
\mathrm{E_{\rm{nanoparticle}}} = \mathrm{E_{\rm{bulk}}n_{\rm{bulk}}} + \sum_{\rm{i}}^{\text{facets}} \mathrm{E_{\rm facet,i}} \mathrm{n_{\rm{facet},i}} + \sum_{\rm{j}}^{\text{defect types}} \mathrm{E_{\rm{defect},j}} \mathrm{n_{\rm{defect},j}}
\end{equation}
where $\mathrm{E_{\rm{bulk}}}$ is the energy per TiO$_2$ unit for the crystal, $\mathrm{n_{\rm{bulk}}}$ is the number of bulk-like TiO$_2$ units in the nanoparticle, $\mathrm{E_{\rm{facet,i}}}$ is the average facet energy per TiO$_2$ unit for the $\rm{i^{th}}$ slab type, $\mathrm{n_{\rm{facet,i}}}$ is  the number of $\mathrm{i^{th}}$ slab-like surface TiO$_2$ units in the nanoparticle, $\mathrm{E_{\rm{defect},j}}$ is the defect energy per TiO$_2$ unit of the $\mathrm{j^{th}}$ defect type, and $\mathrm{n_{\rm{defect,j}}}$ is the number of $\mathrm{j^{th}}$ defect-like TiO$_2$ units in the nanoparticle. We have separated surface-like defects from other defects, since the energy of surface defects can be obtained from slab calculations. In the current work, we consider 3 slab types, i.e., (100)-symmetric, (010)-asymmetric, and (001)-symmetric, and 3 defect types, i.e., edge, corner, and sub-edge. The values for $\mathrm{E_{\rm{defect},j}}$ are determined via least-squares regression, with the values for the remaining quantities computed using the methodology outlined below. 

The energies $\mathrm{E_{\rm{nanoparticle}}}$ and $\mathrm{E_{\rm bulk}}$ are immediately available from DFT calculations for the nanoparticle and bulk, respectively. To calculate $\mathrm{E_{\rm surface,i}}$, we adopt the extrapolation scheme of \citeauthor{Fiorentini_1996} \cite{Fiorentini_1996}. In particular, for each of the slab types, we first calculate the energy of the slab as a function of the number of layers $\mathrm{N}$, which we denote by $\mathrm{E_{\rm slab,i}(N)}$. Next, we determine the average surface energy by fitting the data to the relation:
\begin{equation}
\mathrm{E_{\rm slab,i}(N)} = 2 \mathrm{E_{\rm surface,i}}  + \mathrm{N} \mathrm{\tilde{E}_{\rm bulk,i}} \,
    \label{eq:eq_final_interpolation}
\end{equation}
where $\mathrm{\tilde{E}_{\rm bulk,i}}$ is the extrapolated bulk energy. In the current work, we have used $\mathrm{N=12}$ layers, which results in $\mathrm{E_{\rm surface,i}}$ values converged to within $0.0002$ Ha/atom. Note that for the (010)-asymmetric slab, though the computed $\mathrm{E_{\rm surface,i}}$ is the average over the two different surfaces on either side of the slab, it can be used in our formulation since all nanoparticles considered in this work have both surfaces of the (010)-asymmetric slab, if present at all.

To be able to systematically determine $\mathrm{n_{\rm bulk}}$, $\mathrm{n_{\rm slab,i}}$, and $\mathrm{n_{\rm defect,i}}$, we develop a machine learning scheme that determines the classification of each TiO$_2$ unit in the nanoparticles based on the local atomic environment. Specifically, we use the atomic descriptors proposed by \citeauthor{PhysRevLett.98.146401} \cite{PhysRevLett.98.146401}, details of which can be found in the SI. We generate the atomic descriptors for fingerprinting the model space by using the Atomistic Machine-learning Package (AMP)\cite{Khorshidi2016Amp:Simulations} and select the hyperparameters of the featurization scheme such that there is maximum separation between different geometrical configurations in the model space. We apply dimensionality reduction using kernel principal component analysis (kPCA)\cite{Scholkopf1998NonlinearProblem} on the scaled descriptors to generate linearly independent features and use MeanShift clustering \cite{Anand2014Semi-supervisedClustering}, a density-based clustering algorithm to form clusters of atomic configurations for different types of titanium atoms in nanoparticles. The dimensionally reduced features facilitate visualization in a lower-dimensional space by using the interactive visualization tool ElectroLens.\cite{Lei2019ElectroLens:Features} The scikit-learn software package \cite{Pedregosa2011Scikit-learn:Python} is used for all machine-learning models. We train the clustering algorithm with the model space and use ElectroLens to assign clusters to class labels for the categories of interest. 

\subsection*{Nearsightedness analysis for electronic interactions}
We perform the nearsightedness analysis for electronic interactions using the real-space Spectral Quadrature (SQ) method \cite{suryanarayana2018sqdft, pratapa2016spectral,  suryanarayana2013spectral}, a technique developed for performing large-scale linear-scaling Kohn-Sham DFT calculations. In particular, for the electronic ground state computed using standard diagonalization-based schemes in SPARC \cite{XU2021100709, ghosh2017sparc2, ghosh2017sparc1}, we determine the convergence in electron density with size of interaction region --- quadrature order to be large enough to make associated errors significantly smaller than those considered in this work ---  analogous to previous such results obtained for the energy and atomic forces in bulk aluminum at various temperatures \cite{suryanarayana2017nearsightedness}. Specifically, for a given interaction length scale, we express the electron density at any point in space as a bilinear form in terms of the Hamiltonian, which is then approximated by a Gauss quadrature rule that remains spatially localized to the interaction region by exploiting the locality of electronic interactions in real-space \cite{prodan2005nearsightedness}, i.e, the exponential decay of the density matrix in real-space for insulators as well as metals at finite temperature. Indeed, in the limit of infinite size for the interaction region, the ground state electron density computed using diagonalization is exactly recovered. 


\section{Results and Discussion}

 To study the finite size effects in surface properties and electronic structure of nanoparticles, we construct the ``TiO$_2$ model space'' which is the collection of all model systems using the equilibrium structural parameters. This is illustrated in \cref{fig:model_space}. For the bulk rutile system, we find equilibrium lattice constants of $a = 4.636$  \AA \ and $c = 2.967$ \AA. These lattice constants are in good  agreement with the literature values and are summarized in \cref{supp:tab:lattice_constant} in the SI. The slab models include a mix of symmetric and asymmetric slabs, and the facet energy of asymmetric slabs is calculated using the average between the two different facets. The cuboidal nanoparticles are non-periodic slabs, with vacuum in all directions instead of periodic boundary conditions. These cuboidal isolated nanoparticles are stoichiometric, and the faces of these cuboidal nanoparticles have an atomic structure consistent with the low-index facets of TiO$_2$ slabs. The faces of cuboidal particles will also have defect sites arising due to edges, corners and sub-edge atoms at edges (sub-edges) that are not present in the facets of slab models. These nanoparticles are a convenient way of studying the trends of surface properties and comparing with the extended slab models since their geometries are most similar to the slab models, making it easier to isolate electronic finite size effects. However, the unphysical nature of the cuboidal geometries may cause artifacts when studying system-level properties such as the electronic gap and DOS. Hence, non-stoichiometric nanoparticles that are closer to realistic systems are also created using the Wulff-construction algorithm.\cite{Wulff1901XXV.Krystallflachen} To facilitate comparison between the particle models, we restrict the Wulff-constructed particles to only contain the same low-index facets that appear on the cuboidal particles (100, 010, and 001 facets).

{\begin{figure}[ht!]
    \centering
    \includegraphics[trim=60 20 65 55,clip=True,keepaspectratio=true,scale=0.75]{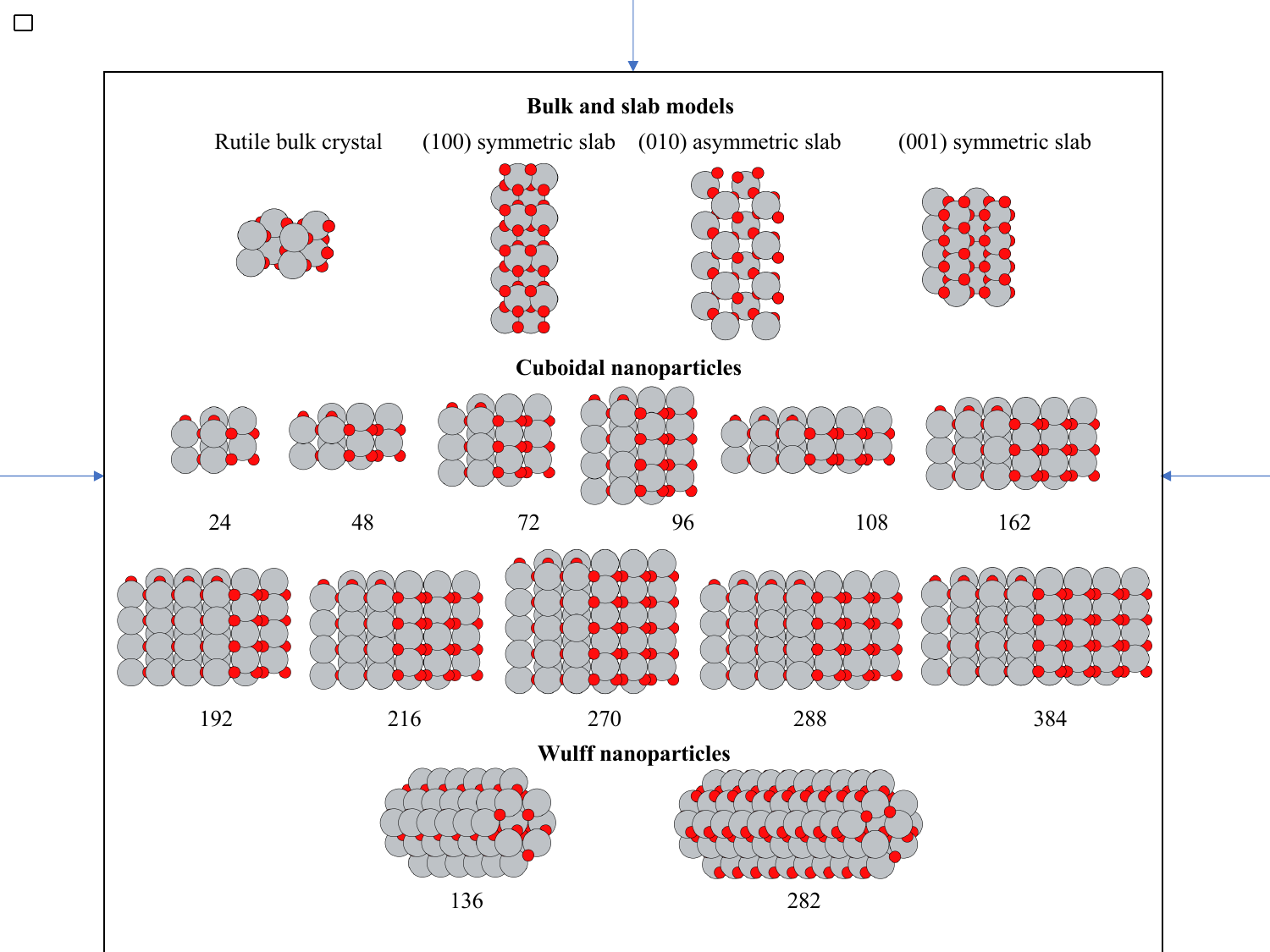}
    \caption{All TiO$_2$ systems studied including the rutile bulk crystal, slab model, cuboidal nanoparticles ranging from 24 to 384 atoms and Wulff nanoparticles constituting of 136 and 282 atoms. Atomic coordinates for all systems are provided in the SI.}\label{fig:model_space}
\end{figure}}

\subsection*{Finite size effects on surface energy}

Finite size effects can be categorized as geometric effects, electronic effects, and quantum effects.\cite{Li2013InvestigationClusters} In the case of energetic properties such as surface or adsorption energies, the geometric finite size effects are expected to be local due to the nearsightedness of electrons.\cite{Prodan2005NearsightednessMatter} Therefore, it is possible to deconvolute the geometric and electronic finite size effects by partitioning the energy to specific types of geometric defects. Any deviation from this partitioning can be assumed to arise from electronic or quantum effects that are implicit in the electronic structure of the particle. Here, we group quantum effects with electronic structure effects since they are both inherent to the behavior of electrons in the system. 

\begin{figure}[ht!]
    \centering
    \includegraphics[clip=True, keepaspectratio=true, scale=0.8]{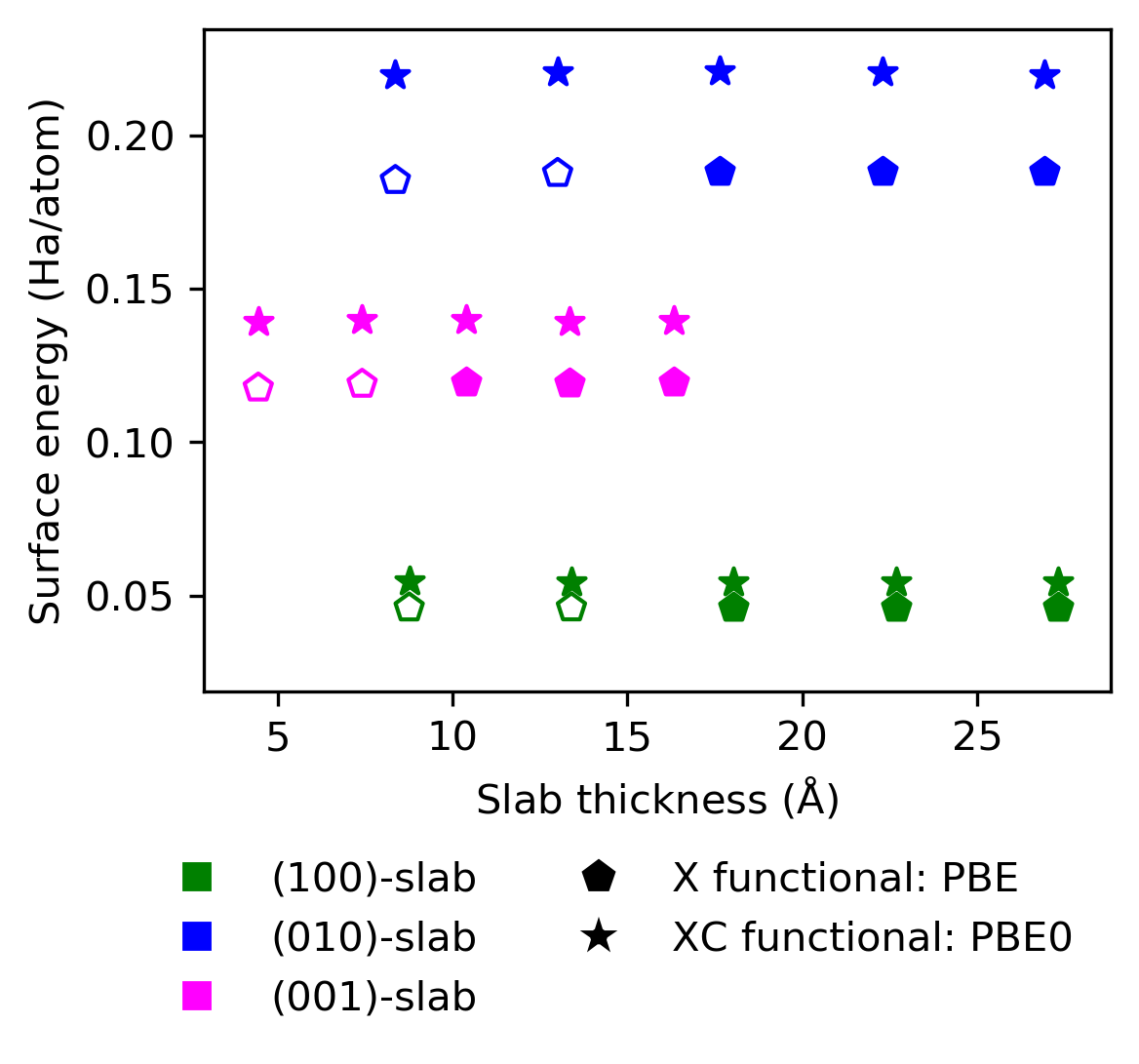}
    \caption{Surface energy (Ha/atom) convergence for TiO$_2$ slabs with varying thickness of slabs from PBE and PBE0 exchange correlation functionals. The hollow symbols represent surface energy from PBE for thin slabs that deviate from the convergent surface energies.}\label{fig:slab_energy_result}
\end{figure}

To quantify the finite size effects on the nanoparticle surface energy, we utilize the total surface energy of the nanoparticle obtained from DFT as the nanoparticle reference:

\begin{equation}
    \mathrm{E_{\rm{surface,DFT}}} =\frac{\mathrm{E_{\rm{nanoparticle}}} - \mathrm{n_{\rm{TiO_{2}}}E_{\rm{bulk}}}}{\mathrm{n_{\rm{surface_{Ti},total}}}}
    \label{eq:ground_truth}
\end{equation}
where $\mathrm{E_{\rm{nanoparticle}}}$ is the total energy of nanoparticle from DFT, $\mathrm{E_{\rm{bulk}}}$ is the bulk energy extracted from the linear interpolation model for slab surface energies (Eq. \ref{eq:eq_final_interpolation}) and $\mathrm{n_{\rm{surface_{Ti},total}}}$ is the total number of TiO$_2$ units belonging to low-index facets and surface defects. In the absence of any finite size effects, the nanoparticle surface energy could be approximated by a linear combination of semi-infinite slab energies:

\begin{equation}
    \mathrm{E_{\rm{surface,slab \ model}}} =\frac{\sum_{\rm{i}}^{\rm{facets}}{\mathrm{E_{\rm{facet, i}}}\mathrm{n_{\rm{facet, i}}}}}{\mathrm{n_{\rm{surface_{Ti},total}}}}
    \label{eq:slab_model}
\end{equation}
where $\mathrm{E_{\rm{facet, i}}}$ is the facet energy and $\mathrm{n_{\rm{facet, i}}}$ is the number of surface-like TiO$_2$ units resembling each facet type. The semi-infinite slab energies are obtained by the linear interpolation method as described in the methods section and are referred to as ``facet energies'' to distinguish them from the energy of other types of surface defects. The converged slab energies are summarized in \cref{tab:surface_en} and are shown in \cref{fig:slab_energy_result}. 


To deconvolute the influence of geometric finite size effects, we utilize a regression model (Eq. \ref{eq:regression}) to obtain the energy contribution of edge, corner, and sub-edge geometric defects present within the system. The total surface energy of the particle based on the regression model can be obtained by adding the energy contributions of all surface defects that constitute of atoms resembling facet types and defects such as corners, edges and sub-edge defects and normalizing to the number of surface Ti atoms:
\begin{equation}
    \mathrm{E_{\rm{surface,regression}}} =\frac{\sum_{i}^{\text{slab types}}{\mathrm{E_{\rm{facet, i}}}\mathrm{n_{\rm{facet, i}}}}+\sum_{\rm{j}}^{\text{defect types}}{\mathrm{E_{\rm{j}}}\mathrm{n_{\rm{defect, j}}}}}{\mathrm{n_{\rm{surface_{Ti},total}}}}
    \label{eq:regression}
\end{equation}
where the term $\sum_{\rm{j}}^{\rm{defect \ types}}{\mathrm{E_{\rm{defect,j}}n_{\rm{defect,j}}}}$ sums over the contributions of corner, edge, and sub-edge Ti atoms and accounts for the geometric finite size effects. The results from the regression model are summarized in \cref{tab:surface_en}.

\begin{table}
  \begin{threeparttable}[b]
  \begin{tabular}{lll}
  \hline
  Defect type & PBE (Ha/atom) & PBE0 (Ha/atom) \\
  \hline
  \hline
  (100)-facet & 0.045 & 0.054\\
  \hline
  (010)-facet & 0.188 & 0.220\\
  \hline
  (001)-facet & 0.119 & 0.139\\
  \hline
  Corner\tnote{1} & 0.188 & 0.223\\
  \hline
  Edge\tnote{1} & 0.151 & 0.197\\
  \hline
  Sub-edge\tnote{1} & 0.099 & 0.121\\
  \hline
  \end{tabular}
  \begin{tablenotes}
    \item[1] Energy contributions of defect sites obtained from regression model. 
  \end{tablenotes}
 \end{threeparttable}
 \caption{Defect energy for different facets and surface defects arising on surface of TiO$_2$ nanoparticles (Ha/atom) from PBE and PBE0 XC functionals.}\label{tab:surface_en}
  \centering
\end{table}

 
\begin{figure}[ht!]
    \centering
    \hspace*{-0.2in}
    \includegraphics[trim=78 95 85 120,clip=True, keepaspectratio=true, scale=0.85]{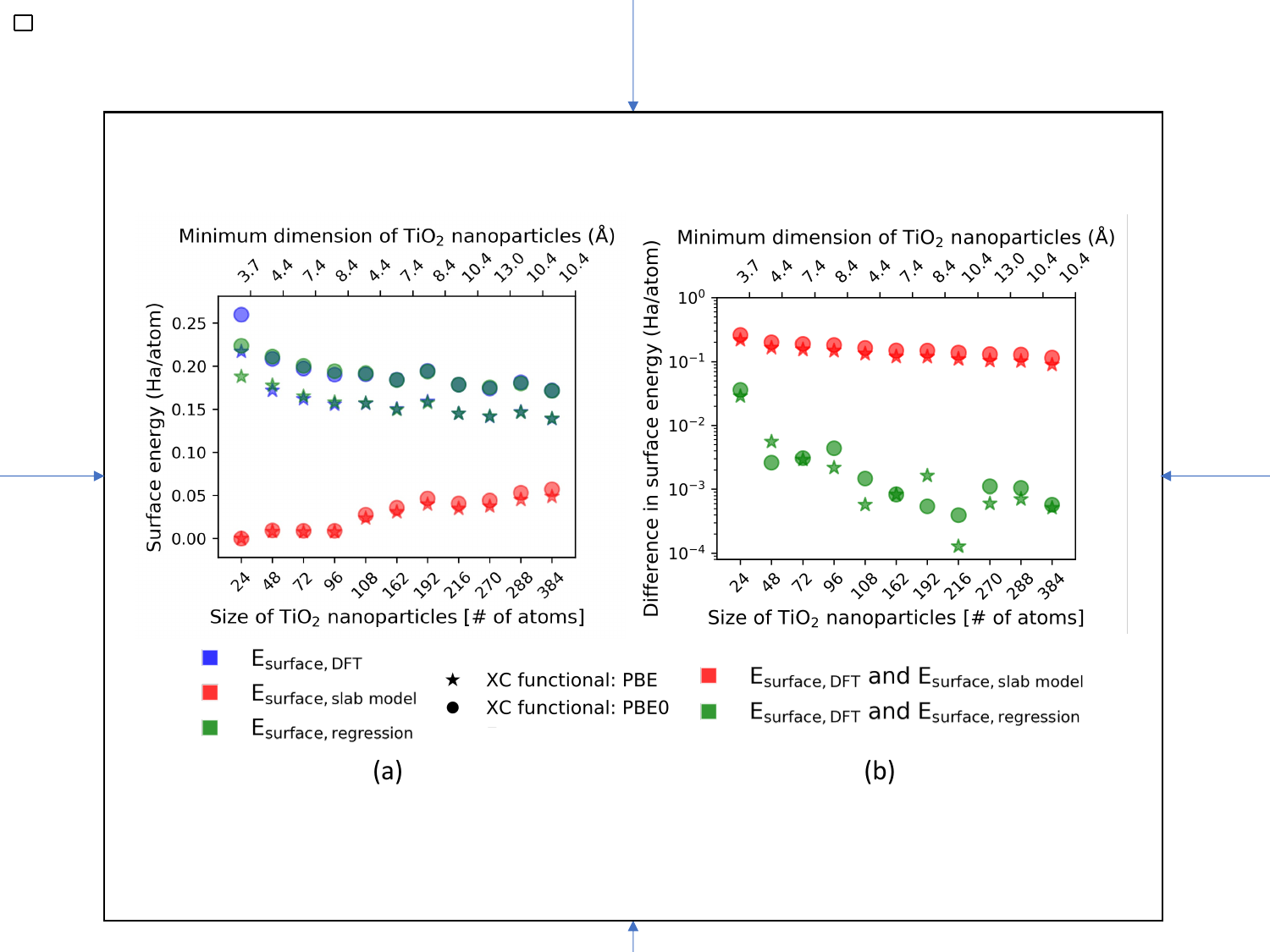}
    \caption{Convergence of surface energy with increasing size of TiO$_2$ nanoparticles for PBE and PBE0 XC functionals. (a) Surface energy from 3 methods: nanoparticle reference DFT ($\mathrm{E_{\rm{surface,DFT}}}$), linear combination of slab models ($\mathrm{E_{\rm{surface,slab \ model}}}$) that does not account for geometric defects and regression model ($\mathrm{E_{\rm{surface,regression}}}$) that accounts for the geometric defects. (b) Absolute difference between $\mathrm{E_{\rm{surface,DFT}}}$ and $\mathrm{E_{\rm{surface,slab \ model}}}$ that quantifies the overall finite size effects
    and absolute difference between $\mathrm{E_{\rm{surface,DFT}}}$ and $\mathrm{E_{\rm{surface,regression}}}$ that quantifies the electronic finite size effects. }\label{fig:nanoparticles_convergence}
\end{figure}

The trends in surface energy convergence for all three models are plotted in \cref{fig:nanoparticles_convergence}. \cref{fig:nanoparticles_convergence}(a) clearly shows that the linear slab model poorly approximates the surface energy, which indicates that finite size effects play a significant role at all particle sizes investigated. However, the regression model provides a much more accurate approximation of the nanoparticle surface energy, indicating that geometric finite size effects are dominant. The residual between the regression model and the nanoparticle reference energy can be interpreted as the contribution due to electronic finite size effects towards the surface energy. The electronic finite size effects are quantified on a log scale in \cref{fig:nanoparticles_convergence}(b). At very small particle sizes ($<$4 \AA{}) the electronic finite size effects are very significant ($\sim$0.03 Ha/atom), which exceeds the typical exchange-correlation error ($\sim$0.01 Ha/atom). At particle sizes from $\sim$4-8 \AA{} ($\sim$30-200 atoms) the electronic finite size effects are below the typical exchange-correlation error, but still exceed the numerical error ($\sim$0.001 Ha/atom). Finally, beyond $\sim$10 \AA{} the electronic finite size effects are reliably at or below the threshold of numerical accuracy, suggesting that they can safely be neglected. This trend is consistent between both PBE and PBE0 functionals. This indicates that the surface properties of TiO$_2$ particles larger than $\sim$ 10 \AA{} can be studied using a combination of slab models, as long as slab models that capture the relevant geometric defects are included.

To validate the regression model, we apply it to particles not included in the regression analysis. These particles are generated by extending four of the cuboidal nanoparticles (24, 72, 108, and 162 atoms) in the 100, 010, and 001-directions and computing their surface energies with PBE. These particles will have the same types of geometric defects (edges, corners, and sub-edge atoms), but the proportions will be different from the original cuboidal particles. The electronic finite size effects (the absolute difference between $\mathrm{E_{\rm{surface,DFT}}}$ and $\mathrm{E_{\rm{surface,regression}}}$) as a function of particle size are shown in the SI (\cref{supp:fig:surface_en_all}). Similar to \cref{fig:nanoparticles_convergence}, the residual decreases to $\sim$1e-3 Ha/atom beyond $\sim$ 10\AA{}, which confirms that the regression model can be generalized to systems that were not included in the training procedure. This also provides further support for the conclusion that the surface properties of nanoparticles with a minimum dimension greater than 10 \AA{} can be accurately modeled using appropriate semi-infinite slabs. 

\subsection*{Characteristic length scale of electronic interactions}

\begin{figure}[ht!]
    \centering
    \includegraphics[clip=True, keepaspectratio=true, scale=0.9]{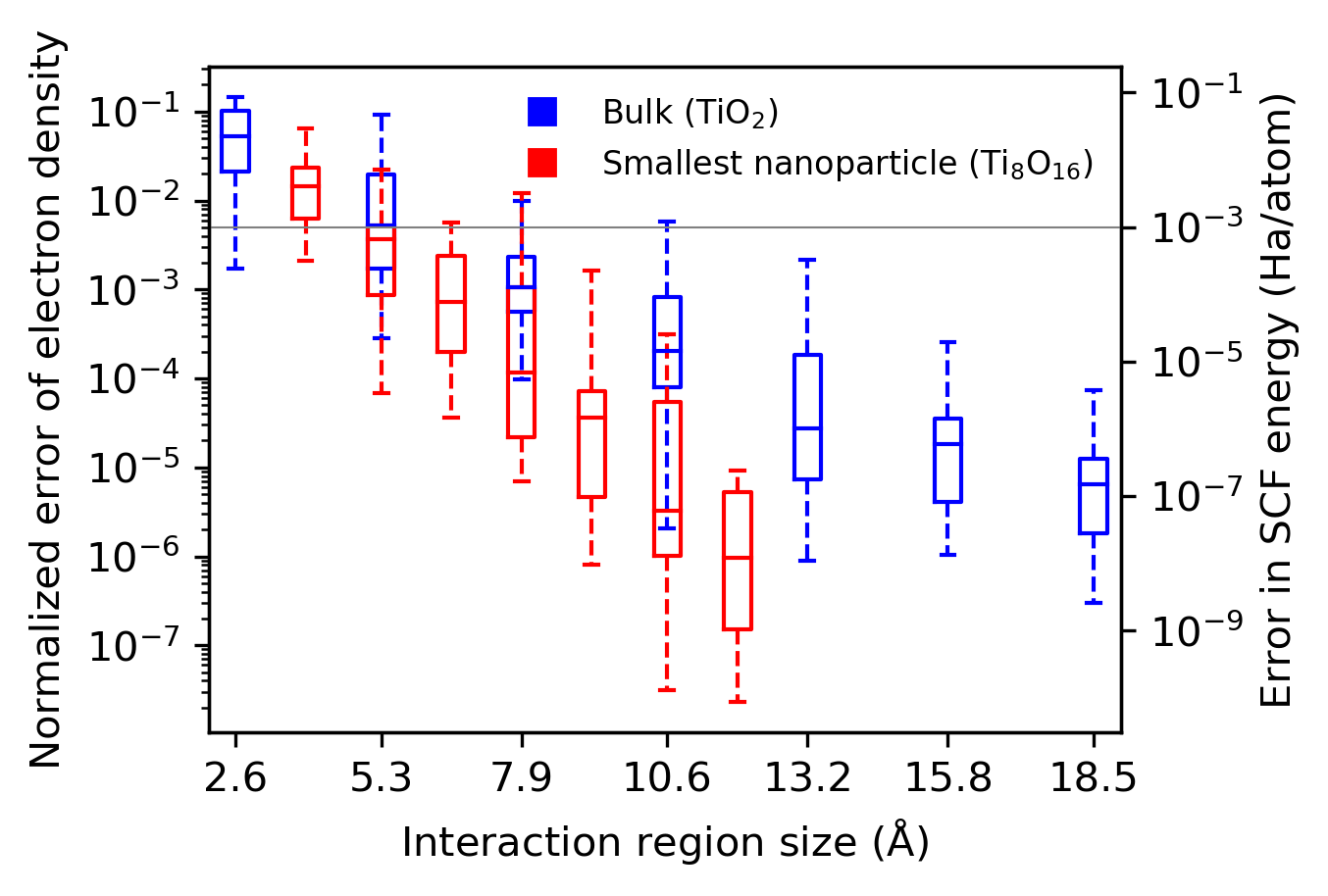}
    \caption{Convergence in electron density with size of interaction region for bulk crystal and the smallest nanoparticle. The error in the self-consistent field (SCF) energy corresponds to the accuracy in the SCF iteration for the lowest normalized error in electron density.(see SI Fig. \cref{supp:fig:callibration_curve}).}\label{fig:nearsight_lin_fit}
\end{figure}


The results of this work indicate that electronic finite size effects have a negligible impact on the surface energetic properties of TiO$_2$ particles larger than $\sim$10 \AA{}. However, this conclusion is material dependent, with the characteristic length scale of finite size effects depending on the localization of electrons in the material. TiO$_2$ is a large bandgap semiconductor, which suggests that electronic finite size effects will decay quickly compared to more delocalized systems such as metals. Indeed, this is consistent with the fact that adsorption energies on metal nanoparticles tend to converge around $\sim$2-3 nm \cite{Li2013InvestigationClusters, Kleis2011FiniteSolids}. Yet, a quantitative assessment of this length scale is extremely computationally demanding with the methods presented here and in prior work, since it requires the calculation of many large particles with DFT.

An alternative approach is to directly characterize the length scale that affects the electronic convergence in the bulk solid. This is achieved by the ``nearsightedness analysis'' discussed in the methods section, which is based on the convergence of the density error as a function of the length scale of the interaction region around each point in the system. The results are shown in \cref{fig:nearsight_lin_fit} for the bulk crystal and smallest nanoparticle. The box plots for the bulk crystal indicate that at a length scale of 10.6 \AA{} the normalized electron density error decays to under 0.01 in all cases and below 0.001 in most cases. To correlate this to energy convergence, the normalized electron density error is compared to the total energy error at a variety of SCF tolerances for bulk TiO$_2$ and a linear fit is used to provide an upper bound based on the systems studied in this work (see SI \cref{supp:fig:callibration_curve}). The results indicate that a normalized electron density error of 0.01 - 0.001 corresponds to an SCF energy error of $\mathrm{10^{-4}-10^{-6}}$ Ha/atom. To ensure that these findings hold for the nanoparticle systems we also perform the nearsightedness analysis for the smallest nanoparticle, Ti$_8$O$_{16}$. The results indicate that the interactions decay as fast or faster in the nanoparticle system, with the error in SCF energy decaying to under 0.001 Ha/atom at a length scale of 8.0 \AA{} in most cases. 

The results of the nearsightedness analysis for both bulk and nanoparticle systems are consistent with the analysis of convergence of surface energy for nanoparticles, where we observe that the electronic finite size effects disappear at a length scale $>$10.0 \AA{}. The nearsightedness analysis and the corresponding energy error calibration can be performed using only the bulk system, and can be applied to any material. This suggests that the nearsightedness analysis is a far more computationally efficient route for evaluating the characteristic length scale of electronic finite size effects for a given material.

\subsection*{Finite size effects on electronic gap and DOS}

The band structure of TiO$_2$ plays an important role in photocatalysis, especially the bandgap and defect states within the gap. To assess the finite size effects on the eigenvalue distribution of the particles, we analyze the electronic gap and DOS of stoichiometric TiO$_2$ nanoparticles and compare the results with bulk and slab models. The experimental bandgap for bulk rutile TiO$_2$ is $\sim$ 3.0-3.1 eV \cite{Bagayoko2011ToBagayoko,Hossain2008OpticalMaterial}, while the bulk gap predicted by PBE and PBE0 are 1.89 and 4.19 eV, respectively. This is consistent with the well-known underestimation of band gaps by PBE,\cite{Borlido2020Exchange-correlationLearning,Tosoni2012electronicTiO2,Mohamad2015APolymorphs} and prior reports of over-estimation of TiO$_2$ bandgaps with hybrid functionals \cite{Fritsch2020,Lee2011InfluenceTiO2}. Nonetheless, the hybrid results are considered to be more reliable since they accurately incorporate exchange interactions that lead to electron localization.

\begin{figure}[ht!]
    \centering
    \hspace*{-0.15in}
    \includegraphics[trim=60 110 70 100,clip=True, keepaspectratio=true, scale=0.75]{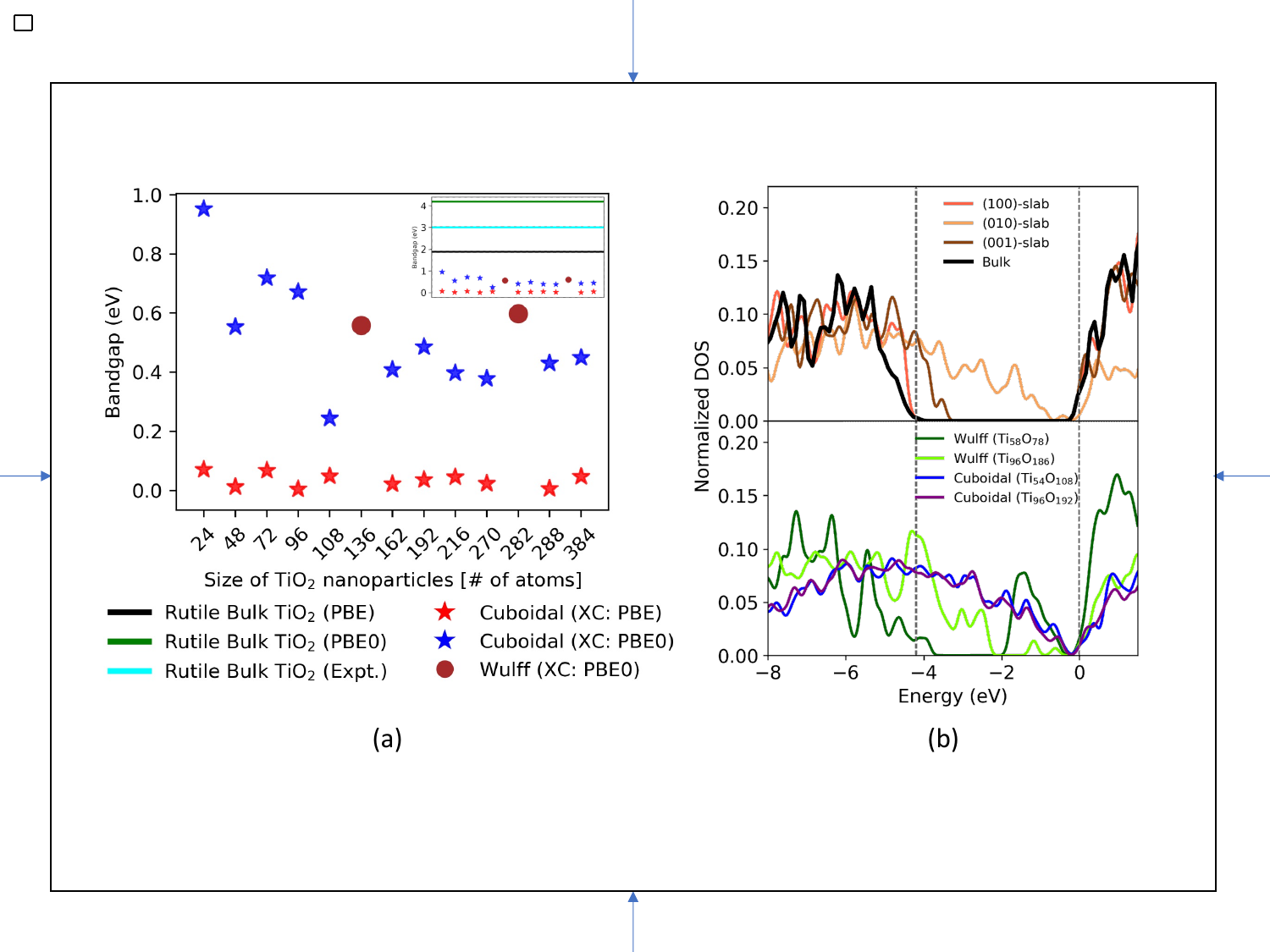}
    \caption{(a) Convergence of electronic gap with increasing size of TiO$_2$ nanoparticles for XC functionals PBE and PBE0. The electronic gaps are compared with the bulk bandgap from 3 methods which are given by 3 horizontal lines in the inset plot: experimental, PBE and PBE0. (b) Comparison of DOS of cuboidal and Wulff-constructed nanoparticles with TiO$2$ bulk and slab models. The top panel shows the DOS of bulk and slab models and highlights the defect states within the gap for (010) and (001)-slab models. The bottom panel depicts the DOS of stoichiometric cuboidal and non-stoichiometric Wulff particles.}\label{fig:bandgap_DOS}
\end{figure}

The electronic gaps for all nanoparticles are shown in \cref{fig:bandgap_DOS}(a) as a function of particle size. The plot indicates that electronic gaps for all nanoparticles are much smaller than the bulk for both PBE and PBE0. Moreover, the electronic gap values decrease, rather than increase as a function of particle size, indicating that the particle sizes investigated are far from the bulk limit. The PBE results suggest that the particles are metallic, indicating a qualitative failure of PBE. For this reason, PBE is omitted from subsequent analyses. 

To understand the reason for smaller electronic gaps in the nanoparticles, we analyze the full DOS of two of the cuboidal particles constituting of 162 and 288 atoms. We compare the DOS of these particles with that of the bulk crystal and slab models and these are plotted in \cref{fig:bandgap_DOS}(b). The DOS is normalized with respect to the total number of electrons in each of the particles. The vertical dotted lines highlight the bandgap of the bulk crystal. In the DOS of cuboidal particles, we observe that the electronic gaps are much lower compared to the bulk bandgap and defect states within the electronic gap are forming a continuum above the HOMO of the bulk crystal. This continuum leads to a very small gap of $\sim$0.2 eV. 

We compare the DOS of the cuboidal particles with the DOS of slab models. We observe that the (100)-slab model has a bandgap very close to the bulk and there are no defect states within the gap, whereas the DOS of the (010) and (001)-slabs reveal the defect states present within the gap resulting in a lower electronic gap. The (010)-facet which is highly unstable also has a continuum of states within the gap, which suggests that the slab models provide some insight into the defect states within the gap for these cuboidal particles. However, these cuboidal particle models are unrealistic due to the presence of high energy surfaces in equal proportions. Therefore, we hypothesize that the nature of DOS of cuboidal particles is an artifact of the unphysical nature of the particles. To  test our hypothesis, we plot the DOS of Wulff-constructed TiO$_2$ particles which are more realistic, since the area occupied by each low-index facet is inversely proportional to the surface energy of the low-index facet. These particles have larger gaps with discrete defect states within the gaps. However, these gaps do not seem to converge to the bulk limit as the particle size increases and the location of the defect states cannot be predicted from the slab DOS plots. 

\begin{figure}[ht!]
    \centering
    \includegraphics[trim=120 25 120 30, clip=True,keepaspectratio=true,scale=0.55]{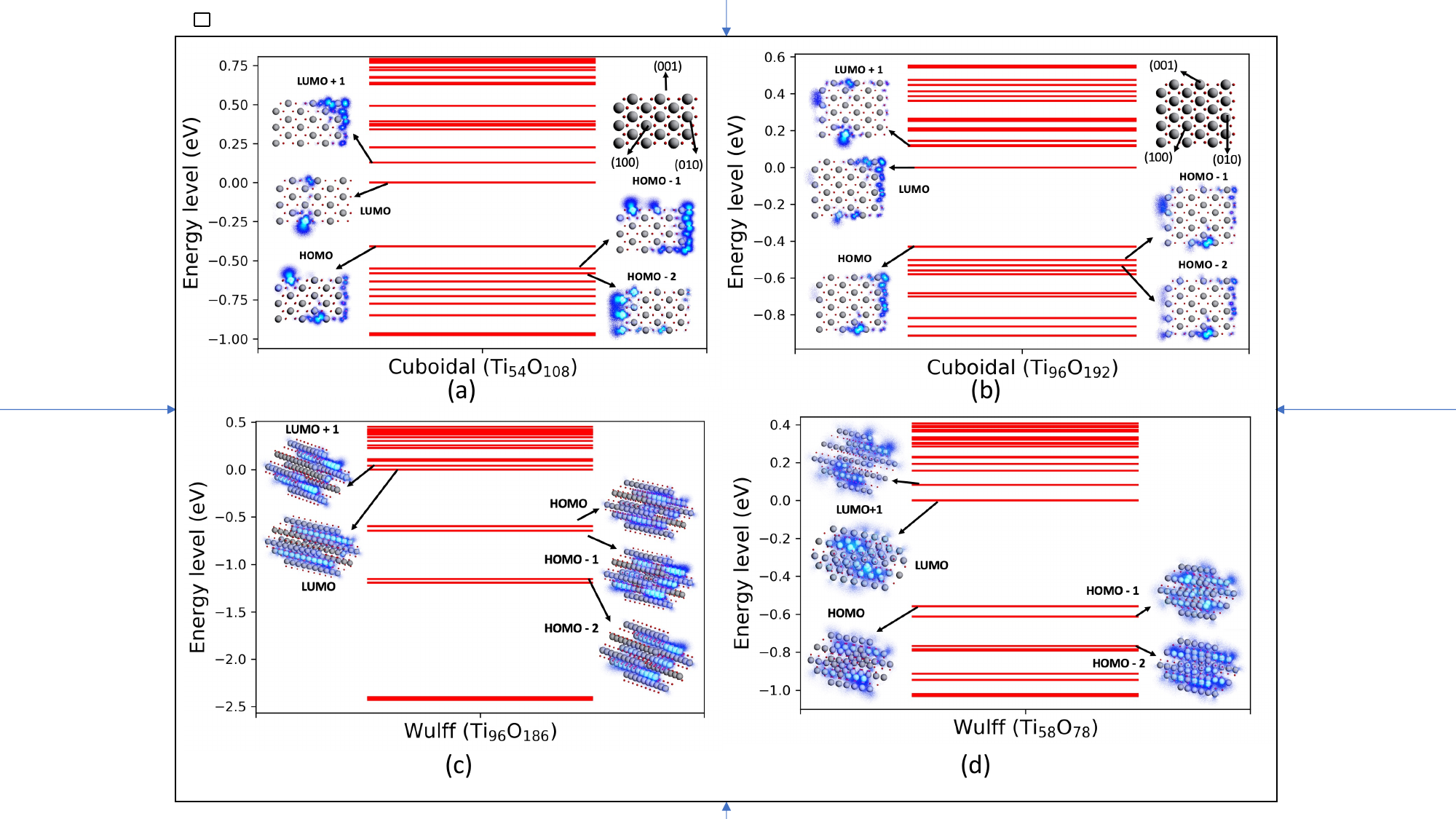}
    \caption{Orbital visualization and  corresponding energy levels for cuboidal (a-b) and Wulff (c-d) nanoparticles. The low-index facets of cuboidal nanoparticles are highlighted in the figure, and Wulff particles are rotated to show states in the bulk.}\label{fig:orbitals_nanoparticles}
\end{figure}

The above results reveal that the electronic gap of Wulff nanoparticles decreases with increase in particle size and is smaller than bulk, in contrast to that  expected from quantum confinement. This observation is consistent with the work of \citeauthor{ko2017size} where it was found that the electronic gap of Wulff-constructed rutile nanoparticles decreased as the particle size decreased.\cite{ko2017size} They suggested that this was due to surface-induced defect states near the HOMO. Visualization of the orbitals for cuboidal particles in \cref{fig:orbitals_nanoparticles} (a-b) reveals that the defect states near both the HOMO and LUMO are highly localized around surface of the particle, especially the (010) and (001) surfaces which have 4-coordinated Ti atoms. This is also consistent with the work of \citeauthor{morales2019understanding} who showed that under-coordinated Ti centers are largely responsible for defect states near the HOMO.\cite{morales2019understanding} However, visualization of the orbitals near the HOMO and LUMO of the Wulff constructed particles in \cref{fig:orbitals_nanoparticles} (c-d) reveals that the defect orbitals in these systems are much more delocalized, and are often contained in the bulk of the particles, particularly for the states near the HOMO. This suggests that a different mechanism is responsible for the decreasing gap in these particles, where there is a complex interaction between surface and bulk that gives rise to electronic defect states that can not be directly assigned to any surface facet or atom. This indicates that hybrid-level calculations of specific geometries of non-equilibrium nanoparticles with sizes up to and beyond 10 \AA{} are required to understand their electronic structure and optical properties.



\section{Concluding Remarks}

In this work, we study the convergence of surface energy and electronic gap as a function of nanoparticle size for rutile TiO$_2$. We utilize the PBE GGA functional and PBE0 hybrid functional, and evaluate cuboidal and Wulff-constructed particles with a maximum size of 384 atoms ($\sim$13 \AA{}). The results indicate that geometric finite size effects play a significant role in the particle surface energy, but that they can be accounted for with a simple linear regression algorithm. In contrast, electronic finite size effects on the surface energy decay rapidly, and can be neglected for particles with a minimum dimension of $>$10 \AA{}. These findings are consistent for both the PBE and PBE0 functionals. Analysis of the nearsightedness of the electronic interactions provides further confirmation of the findings and suggests that the characteristic length scale of electronic interactions for a given material can be computed directly from the bulk electronic structure. This analysis may negate the need for expensive nanoparticle simulations to evaluate the length scale at which finite size effects no longer impact energetic properties.

In the case of electronic gap and DOS, the findings indicate that the nanoparticle systems studied are well below the size at which the electronic gap or DOS converges to the bulk limit. The cuboidal particles are found to have a qualitatively different DOS from bulk TiO$_2$, and it does not change significantly with particle size. This can be explained by the DOS of the slab models, since the high-energy (010) surface slab exhibits a similar DOS to the cuboidal nanoparticles. The Wulff construction is used to generate more realistic particles without highly unstable surfaces, and their DOS are more similar to the bulk, exhibiting a large gap with more discrete defects within the gap. However, the size of the gap and the location of the defects varied considerably with particle size, and the electronic gap decreased with increasing particle size. These results suggest that the electronic gap and DOS of TiO$_2$ nanoparticles are highly sensitive to both the particle size and morphology at the length scales investigated here ($<$15.0 \AA{}). This indicates that hybrid-level simulations of specific TiO$_2$ nanoparticle morphologies are required to elucidate their electronic structures.

The findings have implications for the electronic structure theory of nanoparticle systems and nanoparticle catalysis. The findings demonstrate that the finite-difference SPARC DFT code is capable of performing hybrid-level DFT simulations for large ($>$350 atoms) systems, and that a nearsightedness analysis can be used to rapidly assess the characteristic length scale of electronic interactions. For catalysis, the findings show that the finite size effects of surface properties (e.g. surface energy and adsorption energies) are dominated by geometric defects, and can be simulated using appropriate semi-infinite slab models. On the other hand, in the case of photocatalysis or other applications where the details of the electronic structure are important, explicit nanoparticle models are likely required to describe very small nanoparticles below $\sim$ 20 \AA{}. Further work is necessary to evaluate the electronic gap and DOS of more realistic particle morphologies and elucidate the role of solvents and adsorbates. However, the emergence of highly parallelized hybrid DFT codes like SPARC, along with the increasing prevalence of petascale computing resources, represents promising progress toward evaluating these complex phenomena.


\section{TOC graphic}

{\begin{figure}[H]
    \centering
    \includegraphics[trim=130 140 140 30,clip=True,keepaspectratio=true,scale=0.5]{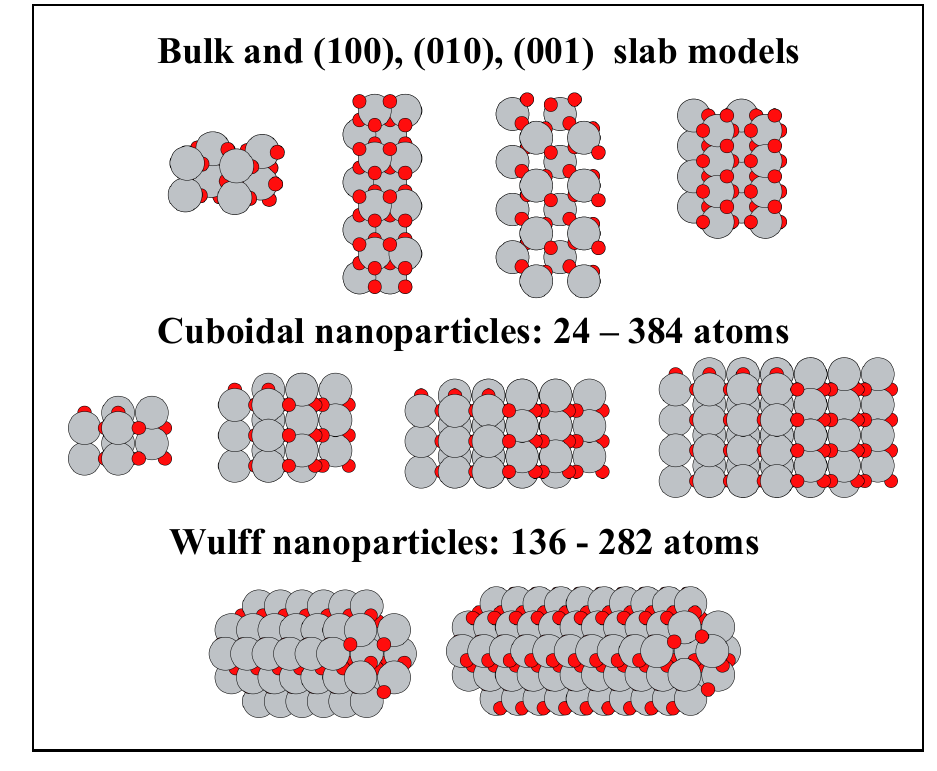}
    \caption*{TOC Graphic}\label{fig:toc_grpahic}
\end{figure}}

\begin{acknowledgement}

The authors acknowledge the funding provided by the U.S. Department of Energy, Basic Energy Sciences, Computational Chemical Sciences Program under grant number DE-SC0019410. This material is also based in part upon work supported by the National Science Foundation under Grant No. 1933646. This research was also supported by the supercomputing infrastructure provided by Partnership for an Advanced Computing Environment (PACE) through its Hive (U.S. National Science Foundation (NSF) through grant MRI-1828187) and Phoenix clusters at Georgia Institute of Technology, Atlanta, Georgia. This report was prepared as an account of work sponsored by an agency of the United States Government. Neither the United States Government nor any agency thereof, nor any of their employees, makes any warranty, express or implied, or assumes any legal liability or responsibility for the accuracy, completeness, or usefulness of any information, apparatus, product, or process disclosed, or represents that its use would not infringe privately owned rights. Reference herein to any specific commercial product, process, or service by trade name, trademark, manufacturer, or otherwise does not necessarily constitute or imply its endorsement, recommendation, or favoring by the United States Government or any agency thereof. The views and opinions of authors expressed herein do not necessarily state or reflect those of the United States Government or any agency thereof.

\end{acknowledgement}

\begin{suppinfo}

The supporting information for this work is available free of charge at \url{https://pubs.acs.org/doi/10.1021/acs.jpcc.1c08915?goto=supporting-info}.\\
The supporting information includes: Equilibrium lattice constants for rutile TiO$_2$ bulk crystal, atom-centered symmetry functions (ACSFs) used for identifying unique atomic configurations in cuboidal nanoparticles, validation of regression model for cuboidal nanoparticles extended in different directions, calibration curve to establish correlation between error in SCF energy and normalized error in electron density, visualization of orbitals around HOMO and LUMO levels of cuboidal and Wulff-constructed nanoparticles, atomic coordinates of TiO$_2$ model systems.
\end{suppinfo}

\bibliography{references_abv}


\end{document}



\thispagestyle{fancy}
\fancyhf{}
\cfoot{S\thepage}
\myexternaldocument{main}

\subsection{Equilibrium lattice constants for rutile TiO$_2$ bulk crystal}

\begin{table}[ht!]
\begin{center}
\begin{tabular}{ c c c }
  & $a\ $(\AA{}) & $c\ $(\AA{}) \\ 
  \hline
  \hline
 Experimental\cite{Montanari2002LatticeCalculations,Moellmann2012AModifications} & 4.587 & 2.954 \\ \hline
 \citeauthor{Montanari2002LatticeCalculations}\cite{Montanari2002LatticeCalculations} & 4.641 & 2.966 \\
 \hline
 \citeauthor{Labat2007DensityFunctionals}\cite{Labat2007DensityFunctionals} & 4.653 & 2.975\\
 \hline
 This work & 4.636 & 2.967
\end{tabular}
\caption{Comparison of equilibrium lattice parameters for bulk rutile TiO$_2$ crystal from experimental and theoretical studies calculated using PBE exchange correlation functional.}\label{tab:lattice_constant}
\end{center}
\end{table}
\subsection{Atom-Centered Symmetry Functions(ACSFs)}
We use the Gaussian descriptors proposed by \citeauthor{PhysRevLett.98.146401}\cite{PhysRevLett.98.146401} which are also known as ACSFs to identify the unique atomic configurations belonging to corner, edge and sub-surface defect sites that are not present in the bulk and slab models of TiO$_2$. The descriptors represent local atomic environments for every atom with respect to neighboring atoms within a given cutoff radius, $R_c$. This fingerprinting scheme generates feature vectors for every atom in the system when provided with atomic positions as an input. This transformation is given by:

\begin{equation}
    \tilde{\textbf{R}} \xrightarrow[\text{feature mapping}]{\text{ACSFs}}  \textbf{G} (\tilde{\textbf{R}}, R_{c}, \vec{W})
    \label{eq:feature_mapping}
\end{equation}

where $\textbf{G}$ is the set of descriptors generated by the fingerprinting scheme. $\vec{W}$ is the set of parameters for radial and angular symmetry functions. The radial and angular symmetry functions are referred to as $G_2$ and $G_4$ respectively. The radial symmetry functions are sum of Gaussians that are multiplied to a cutoff function $f_c(R_{ij})$:

\begin{equation}
    G_{i}^{2} = \sum_{j = 1}^{N} e^{-\eta \left( R_{ij} - R_s \right)^2} \cdot f_c(R_{ij})
    \label{eq:g2}
\end{equation}

The angular symmetry functions are built from angle centered at every atom i for three-body interaction, $\theta_{ijk}$ with neighbor atoms, $j$ and $k$ and is given by the following equation:

\begin{equation}
    G_{i}^{4} = 2^{1-\zeta} \sum_{j \neq i}^{N} \sum_{k \neq i, j}^{N} \left[ \left(  1 + \lambda \cdot \cos{\theta_{ijk}} \right)^{\zeta} \cdot e^{-\eta \left( R_{ij}^2 + R_{ik}^2 \right)} \cdot f_c(R_{ij}) \cdot f_c(R_{ik}) \right]
    \label{eq:g4}
\end{equation}




The central cut-off function is represented by $f_c(R_{ij})$ and depends on the cutoff radius, $R_c$. It captures the inter-atomic interactions between any atom and its neighboring atoms.

The Gaussian fingerprinting of atomic environments has been implemented in the software package, Atomistic Machine-Learning Package (AMP).\cite{Khorshidi2016Amp:Simulations} We use this package to construct feature vectors for the systems of interest in this study. 


\subsection*{Validation of regression model}

\begin{figure}[ht!]
    \centering
    \includegraphics[scale=0.8]{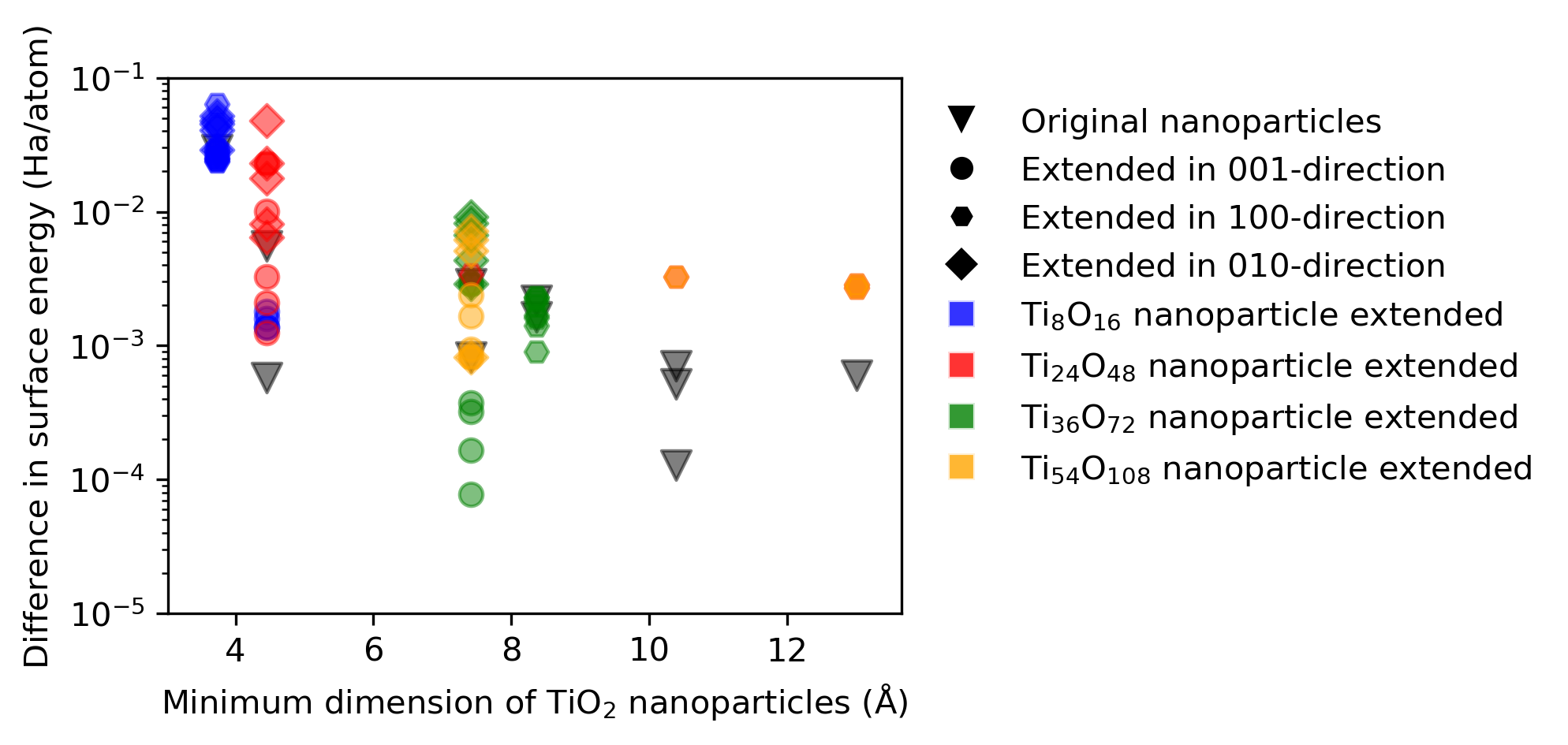}
    \caption{Model validation for regression model quantifying the electronic finite size effects when the nanoparticles containing 24, 72, 108 and 162 atoms are stretched along 3 directions (100), (010) and (001). The difference in surface energy converges between 1e-3 and 1e-2 Ha/atom at a length scale of $\sim$ 12.0 \AA{}. }\label{fig:surface_en_all}
\end{figure}

\newpage
\subsection*{Nearsightedness analysis}
\begin{figure}[ht!]
    \centering
    \includegraphics[clip=True, keepaspectratio=true, scale=0.8]{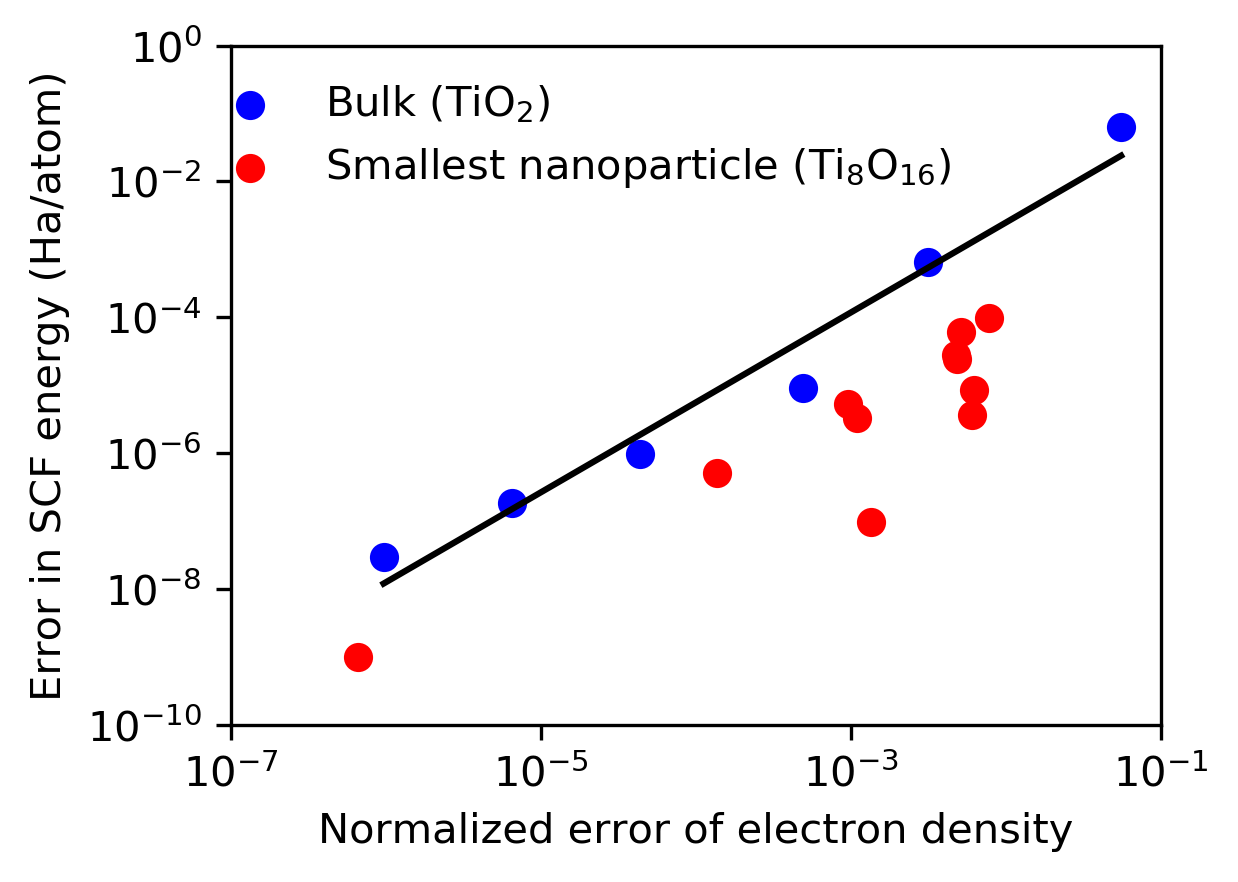}
    \caption{Calibration curve for nearsightedness analysis which establishes the upper bound for correlation between error in SCF energy and normalized error of electron density.}\label{fig:callibration_curve}
\end{figure} 

\newpage
\subsection*{Orbitals of medium-sized cuboidal and Wulff nanoparticles}
\begin{figure}[ht!]
    \centering
    \hspace*{-0.4in}
    \includegraphics[trim=140 80 160 90,clip=True, keepaspectratio=true, scale=0.9]{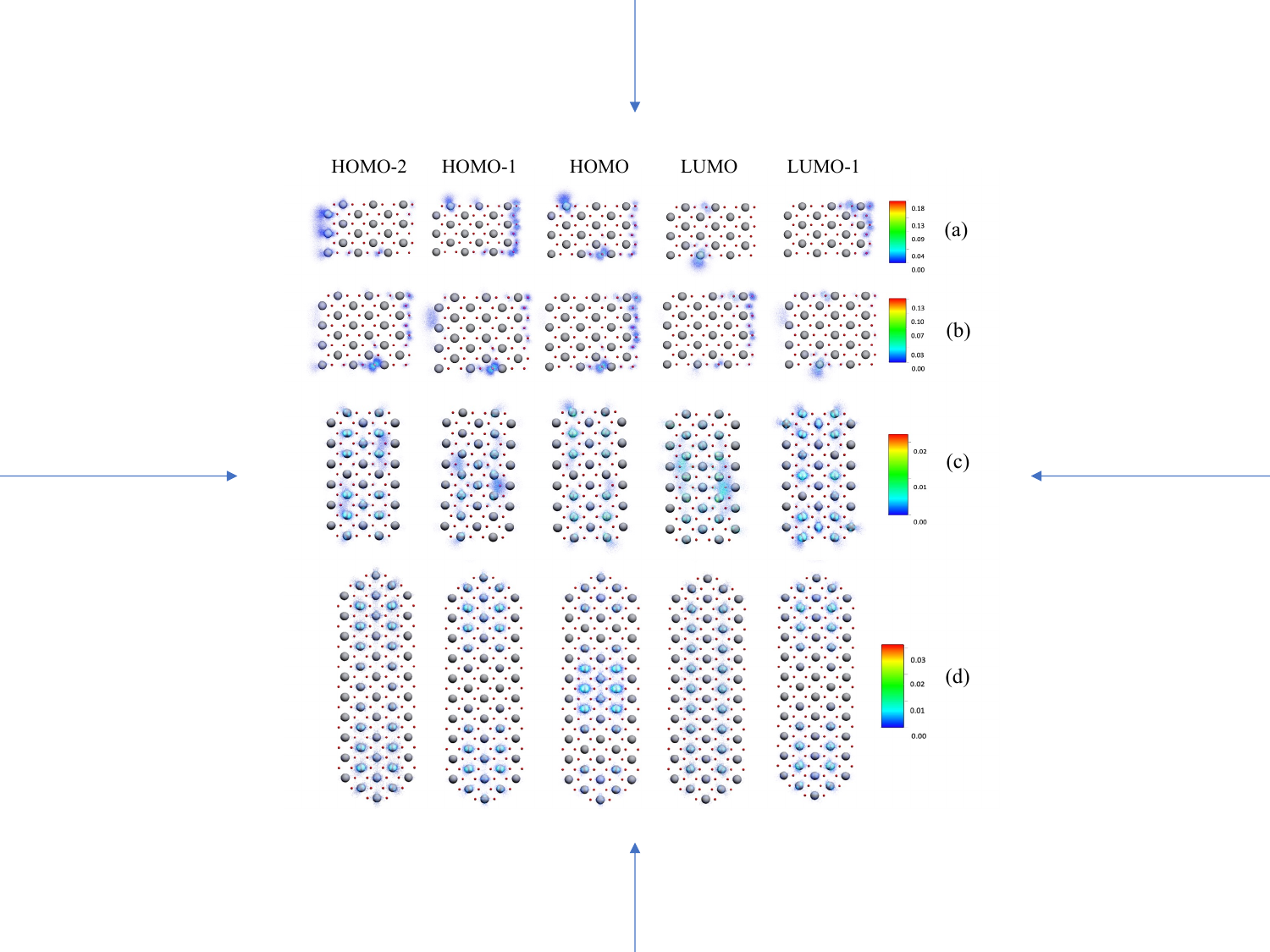}
    \caption{Visualization of orbitals around HOMO and LUMO levels of cuboidal and Wulff-constructed particles using ELectroLens.\cite{Lei2019ElectroLens:Features} HOMO-2 and HOMO-1 refer to the two orbitals just below the HOMO and LUMO-1 refers to the orbital just above the LUMO. These figures indicate that the one-electron densities are localized around the high-energy facets for cuboidal nanoparticles whereas they are more delocalized in the case of Wulff particles.}\label{fig:orbitals}
\end{figure}

\subsection{Atomic coordinates of TiO$_2$ model systems}

Atomic coordinates for rutile TiO$_2$ bulk, slabs, cuboidal and Wulff nanoparticles are available in \texttt{.xyz} format in the file \texttt{atomic\_coordinates.zip}.

\newpage
\bibliography{references}